\documentstyle[epsf,12pt,aps]{revtex}
\begin{document}

\title{Escape from intermittent repellers-\\
Periodic orbit theory for crossover from exponential to algebraic decay}
\author{Per Dahlqvist \\
Mechanics Department \\
Royal Institute of Technology, S-100 44 Stockholm, Sweden}
\maketitle

\vspace{2cm}

\begin{abstract}
We apply periodic orbit theory to study the asymptotic 
distribution of escape times from an 
intermittent map.
The dynamical zeta function exhibits a branch point which is associated with
an asymptotic power law escape.
By an analytic continuation technique we compute a zero of the
zeta function beyond its radius of convergence leading to a 
pre-asymptotic
exponential decay. The time of crossover from an exponential to a power law is 
also predicted.
The theoretical predictions are confirmed by numerical simulation.
Applications to conductance fluctuations in quantum dots
are discussed.
\end{abstract}

\vspace{2cm}

\section{Introduction}

Exponential distribution of escape times
from chaotic scattering systems should be expected only
if the associated repeller is hyperbolic.
For intermittent repellers one expects asymptotic power law decay\cite{Ott,Alt}.
Nevertheless, in numerical simulations one often observes what appears
to be a perfect exponential\cite{jal90,BB90}, and an expected 
crossover to a power law may be hard
to detect, because it may occur 
after a long time
where it is difficult to obtain descent statistics.

The importance of intermittency cannot be overemphasized.
A generic Hamiltonian system exhibits a mixed
phase space structure. A typical trajectory is intermittently trapped close to
the stable islands \cite{perc}. 
But even fully chaotic billiards
may exhibit intermittency, typically if they have neutrally stable
orbits. Popular billiards such as the Stadium and the Sinai billiards
are of this type.

A quantum dot is an open scattering system in two dimensions,
obtained by connecting leads to a cavity. Inspired by Quantum Chaos research,
one likes to contrast shapes of the cavity corresponding to chaotic
motion, like the stadium, with shapes corresponding to integrable motion,
like the rectangle or the square.
Both extreme cases are sensitive to naturally occurring imperfections
and one naturally ends up with mixed phase space systems
where one component hopefully dominates.
Consequently, 
the signals of underlying chaos or integrability don't show up as
clear cut as one would have hoped for.

Much of the analysis of these problems has been numerical and
heuristic.
To strengthen the theoretical analysis we will,
in this paper, apply periodic orbit theory and cycle expansions 
to make quantitative predictions concerning the
asymptotic distribution of escape times from an intermittent map.
We will demonstrate that a {\em pre-asymptotic} exponential  escape
law is associated with a pair of complex conjugate zeroes 
of the zeta function beyond
its domain of convergence. This zero will be computed with a simple
resummation technique\cite{PDJPAL97}.
The truly asymptotic escape distribution will be a power law,
and
is associated with a branch point of the zeta function.
The strength of this power law will also be provided by the resummation scheme, 
whereas
the particular power is known from analytic argument for our particular
model system.
The relative magnitudes of the pre-exponential and the power law will
yield a good estimate of the cross over time, which will be surprisingly
high.

\section{Escape and periodic orbits}

Much of the early work on cycle expansion \cite{AAC} was concerned with escape
from (hyperbolic) repellers, so we can follow ref. \cite{AAC} rather closely
when deriving the basic formulas relating
escape to the  periodic orbits
of the repeller.

Consider a 1-d map, on some interval
$I$,
with $L$ monotone
branches $f_i(x)$ where $1\leq i\leq L$. Each branch $f_i(x)$ is
defined on an interval
$I_i$.
A generating partition  is then given by
${\cal C}^{(1)}=\{ I_1,I_2 \ldots I_L\}$.
We want the map to admit  an unrestricted symbolic
dynamics.
We therefore require all branches to map their domain $f_i(I_i)=I$
onto some interval
$I\supset {\cal C}^{(1)}$ covering ${\cal C}^{(1)}$.
A trajectory is considered to escape when 
some iterate of the map $x\notin {\cal C}^{(1)}$.

\begin{figure}
\centering
\leavevmode
\epsfxsize=11cm
\epsfbox{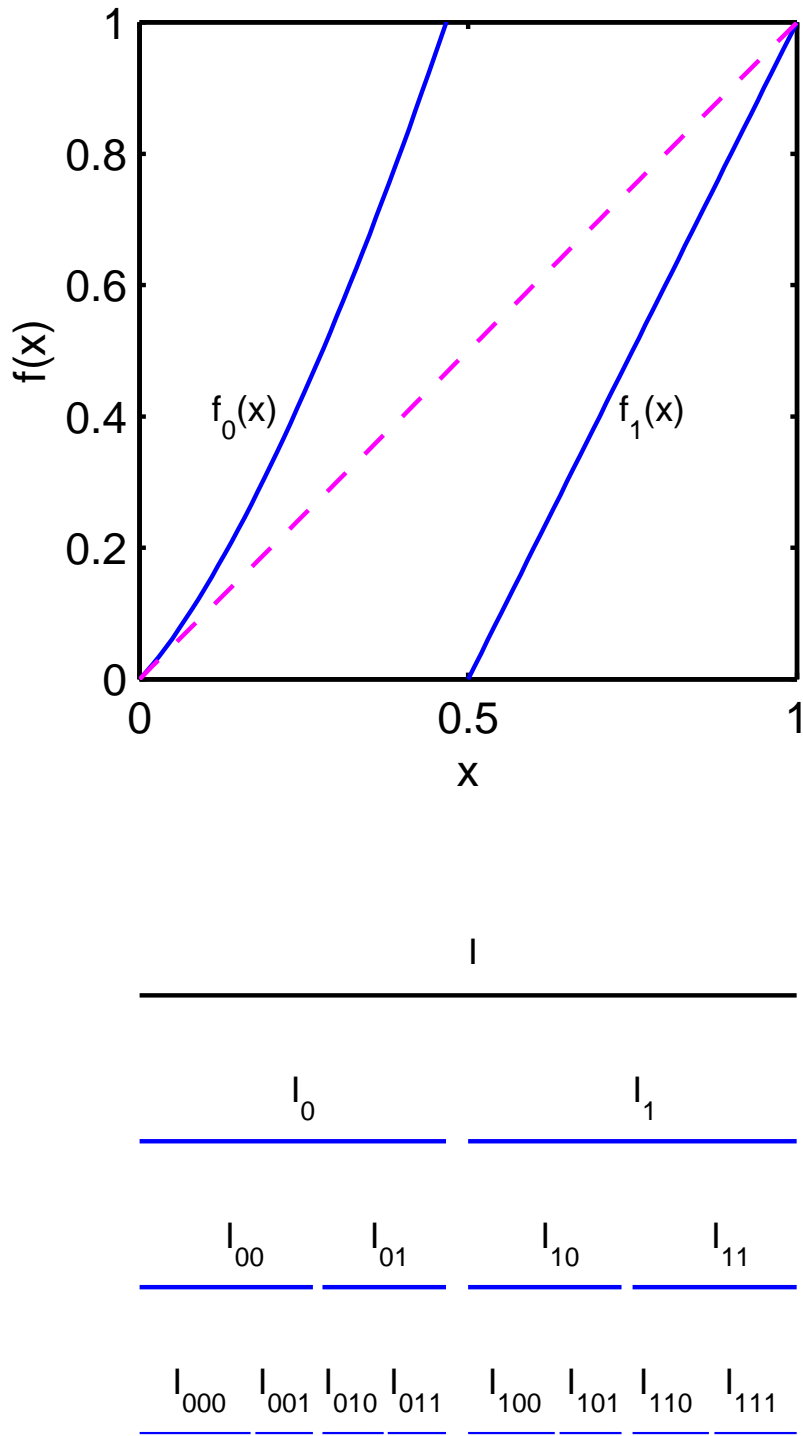}
%
%

\vspace{0.5cm}

\caption{The intermittent map (\ref{eqn:themap}) for the parameter
values $s=0.7$ and $p=1.2$.
The map is defined on the interval $I$. Below the map is also shown the
partitions ${\cal C}^{(1)}=\{ I_0 , I_1 \}$, ${\cal C}^{(2)}$ and 
${\cal C}^{(3)}$.
}
\label{fig:themap}
\end{figure}

The $n$'th level partition ${\cal C}^{(n)}=\{ I_q ; n_q=n\}$
can be constructed iteratively. Here $q$ are words of length n
built from the alphabet ${\cal A}=\{i;0\leq i \leq L\}$.
An interval is thus defined recursively according to
\begin{eqnarray}
I_{iq}=f_i^{-1}(I_q)  \   \    ,
\label{eqn:recrule}
\end{eqnarray}
where $iq$ is the concatenation of letter $i$ with word $q$.
A concrete example will be given in eq. (\ref{eqn:themap}) and fig.
\ref{fig:themap}.
Next define the characteristic function for
the $n$'th level partition
\begin{equation}
\chi^{(n)}(x)=\sum_{q}^{(n)} \chi_q(x)  \    \    ,
\end{equation}
where
\begin{equation}
\chi_q(x)=\left\{ \begin{array}{cc}
1 & x \in I_q\\
0 & x \notin I_q
\end{array}\right.    \ \   .
\end{equation}
An initial point surviving $n$ iterations must be contained in ${\cal C}^{(n)}$.
Starting from an initial (normalized) distribution 
$\rho_0(x)$ we can express the
fraction that survives $n$ iterations as
\begin{equation}
\Gamma_n=\int \rho_0(x)  \chi^{(n)}(x) dx   \  \  .
\end{equation}
We choose the distribution $\rho_0(x)$ to  be uniform
on the interval $I$. The survival probability is then given by
\begin{equation}
\Gamma_n= a \sum_q^{(n)} |I_q|  \  \   ,
\end{equation}
where
\begin{equation}
a^{-1}=\int_{I} dx =|I| \  \  .
\end{equation}

Assuming hyperbolicity  the size of $I_q$ can be related to the
stability $\Lambda_q=\frac{d}{dx}f^n(x) \mid_{x\in q}$ of periodic
orbit $\overline{q}$ according to
\begin{equation}
|I_q| = b_q \frac{1}{|\Lambda_q|}
\label{eqn:absIq}
\end{equation}
where $b_q=O(|I|)$, can be bounded close to the size of $I$.
This results from the fact that
$f^n(I_q)=I$,
the smallness of $|I_q|$
and the fact that 
derivatives
can be bounded due to hyperbolicity.
We will eventually relax the assumption of hyperbolicity,
but for the moment we'll stick to it.

The survival fraction can now be bounded by the periodic orbit sum
according to
\begin{equation}
C_1(N)\; \sum_q^{(n)} \frac{1}{|\Lambda_q|} < 
\Gamma_n <C_2(N) \; \sum_q^{(n)} \frac{1}{|\Lambda_q|}
\label{eqn:boundhyper}
\end{equation}
for all $n>N$.
For large $N$ (and assuming hyperbolicity) $C_1(N)$ and 
$C_2(N)$ can be chosen close to unity.

The periodic orbit sum in (\ref{eqn:boundhyper}) 
will be denoted $Z_n$ 
\begin{equation}
\sum_q^{(n)} \frac{1}{|\Lambda_q|}\equiv Z_n \ \  .
\end{equation}
and can be rewritten as a sum over primitive periodic orbits (period $n_p$) and
their repetitions
\begin{equation}
Z_n =
\sum_p n_p \sum_{r=1}^{\infty} 
\frac{\delta_{n,r\;n_p}}{|\Lambda_p|^r}  \  \  .
\end{equation}

It is closely related to the trace of
the Perron-Frobenius operator
\begin{equation}
Z_n \approx \mbox{tr} {\cal L}^n =\int dx \delta (x-f^n(x))=
\sum_p n_p \sum_{r=1}^{\infty} 
\frac{\delta_{n,r\;n_p}}{|\Lambda^r_p-1|}  \  \  .
\end{equation}

\newpage

By introducing the zeta function
\begin{equation}
\zeta^{-1}(z)=\prod_p \left( 1-\frac{z^{n_p}}{|\Lambda_p|}\right)  \  \  ,
\label{eqn:zzdef}
\end{equation}
the periodic orbit sum $Z_n$ can be expressed as a contour integral
\begin{equation}
Z_n=\frac{1}{2\pi i} \int_\gamma z^{-n} 
\left( \frac{d}{dz} \log \zeta^{-1}(z) \right) dz  \  \  ,
\label{eqn:prev}
\end{equation}
where the small contour $\gamma$ encircles the origin in negative direction.

The expansion of the zeta function to a power series is usually referred
to as a {\em cycle expansion}.
\begin{equation}
\zeta^{-1}(z)=\sum c_n z^n \label{eqn:cycexp}  \  \  .
\end{equation}
This representation converges up to the leading singularity, unlike
the  product representation (\ref{eqn:zzdef})
which diverges at (nontrivial) zeroes.
If the zeta function $\zeta^{-1}(z)$ is analytic in a disk extending beyond
the leading zero $z_0$, then the periodic orbit sum $Z_n$,
and hence the survival probability $\Gamma_n$,
will decay
asymptotically as
\begin{equation}
Z_n \sim z_0^{-n} \equiv e^{-\kappa n}
\end{equation}
where $\kappa = \log z_0$ is the escape rate.

We will introduce intermittency in connection with a specific model.
We then consider  an intermittent map
$x \mapsto f(x)$ with two branches $L=2$, and where $I$ is chosen
as the unit interval: 
\begin{equation}
f(x)=\left\{ 
\begin{tabular}{lll}
$f_0(x)=x\left(1+p \; (2x)^s \right)$ & $x\in I_0=\{x; 0\leq x <\xi\}$\\
$f_1(x)= 2x-1$ & $ x\in I_1=\{x; 1/2 \leq x \leq 1\}$
\end{tabular}  \ \ ,
\right.  \label{eqn:themap}
\end{equation}
The map is intermittent if $s>0$ and allows escape if $p>1$.
The map is shown in
fig. \ref{fig:themap}, together with some of its partitions. 
The right edge of the left branch $I_0$, denoted $\xi$, is implicitly
defined 
by $f_0(\xi)=1$.
The particle is considered to escape if
$\xi(s) < x < \frac{1}{2}$.

The intermittent property is related to the fact that the cycle
$\overline{0}$ is neutrally stable $f'(0)=1$. Consequently,
cycle stabilities can no longer be exponentially bounded with length.
This loss of hyperbolicity makes it difficult
to relate the survival probability $\Gamma_n$
to the periodic orbit sums $Z_n$.
Indeed, eq. (\ref{eqn:absIq}) is brutally violated in some cases,
as can been realized from the following example.
The problem of intermittency is best represented by the family of periodic
orbits $10^k$.
It follows from eqs. (\ref{eqn:recrule}) and (\ref{eqn:themap}) that
$|I_{10^k} |=\frac{1}{2} |I_{0^k} |$. It can be shown \cite{PDJPAL97}
that
$|I_{0^k}| \sim  \frac{1}{k^{1/s}}$ and thus
\begin{equation}
|I_{10^k} | \sim \frac{1}{k^{1/s}}
\end{equation}
which should be compared with the asymptotic behavior of the stabilities
\begin{equation}
1/\Lambda_{10^k} \sim \frac{1}{k^{1+1/s}}
\end{equation}

\begin{figure}
\centering
\leavevmode
\epsfxsize=10cm
\epsfbox{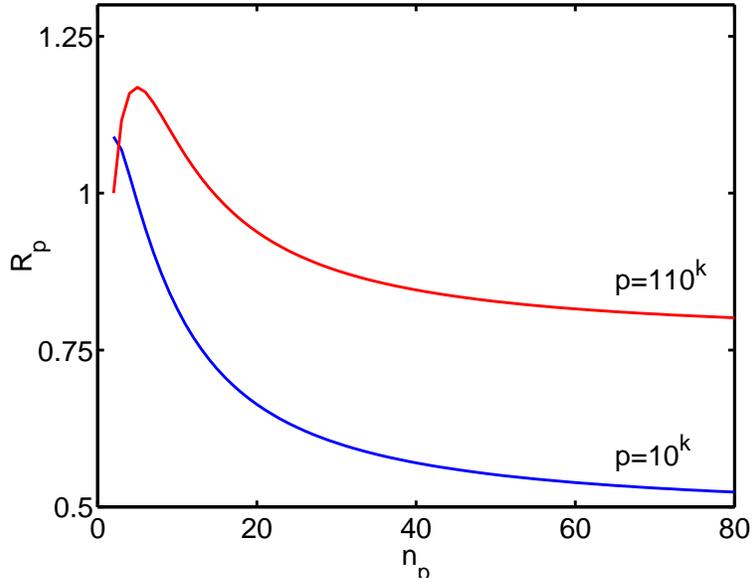}
\caption{The quantity $R_p$ plotted for the sequences
$p=10^k$ and $p=110^k$ versus length $n_p$.}
\label{fig:bounds}
\end{figure}

The difference in power laws seems to spoil every possibility
of a bound like (\ref{eqn:boundhyper}).
However, eq. (\ref{eqn:absIq}) is not  necessary 
for that purpose.
It suffices if the ratio
\begin{equation}
R_p= \frac{|\Lambda_p|}{n_p} \sum_{k=1}^{n_p} |I_{{\cal S}^k p}|
\end{equation}
stays bounded. 
That it to say that it suffices if the average size of the intervals along
a cycle can be related to the stability, rather than each interval separately.
Here ${\cal S}$ denotes the shift operator.
We check this numerically on two sequences: $10^k$ and $110^k$,
the former being
most sensitive to intermittency. 
The result is plotted in fig.
\ref{fig:bounds}. 
We note that for both sequences, $R_p$ appear to tend to well defined limits,
where $10^k$ lead to the largest deviation from unity.
Indeed,
it is reasonable to assume that the sequence $10^k$
provides a lower bound
\begin{equation}
R_p > \lim_{k\rightarrow \infty} R_{10^k} \hspace{0.8cm} \forall p \  \  .
\end{equation}
In view of this, the
numerical results strongly suggest that
$R_p$ stays bounded, and that, for this particular system, 
$C_1$ in eq. (\ref{eqn:boundhyper}) can be chosen as $C_1=0.5$ and 
$C_2$ presumably close to unity.
This is a surprisingly low price to pay for the complication of
intermittency.

The sizes of the intervals
$I_{0^n}$ has no relation whatsoever to the stability of the
cycle $\overline{0}$, which is unity.
We exclude the intervals
$I_{0^n}$
from our considerations by
{\em pruning} the fixpoint from the zeta function
\begin{equation}
\zeta^{-1}(z)=\prod_{p\neq 0} \left( 1-\frac{z^{n_p}}{|\Lambda_p|}\right)  \  \  ,
\label{eqn:zzdefprun}
\end{equation}
The contribution from $I_{0^n}$ to $\Gamma_n$ can be added
separately if required.

Since the result rely on summation along periodic orbits, it might break down for
some choices of the initial distribution $\rho_0(x)$  were such a summation is
not carried out.

\section{Resummation and simulation}

After having argued that the survival probability $\Gamma_n$ 
still can be bounded
close to periodic orbit sums $z_n$  we turn to the problem of computing the
asymptotics of these periodic orbit sums.
The coefficients of the cycle expansion 
(\ref{eqn:cycexp}) for the map (\ref{eqn:themap}) decay asymptotically as
\begin{equation}
c_n \sim \frac{1}{n^{1+1/s}}   \  \  ,
\end{equation}
which induces a singularity of the 
type $(1-z)^{1/s}$ in the zeta function \cite{PDJPAL97}.
If $1/s$ is an integer, the singularity is
$(1-z)^{1/s}\log (1-z)$.

To evaluate the periodic orbit sum it is convenient to consider 
a resummation of the zeta function 
around the branch point $z=1$.
\begin{equation}
\zeta^{-1}(z)=\sum_{i=0}^{\infty} c_i z^i =
\sum_{i=0}^{\infty} a_i (1-z)^i+
(1-z)^{1/s} \sum_{i=0}^{\infty} b_i (1-z)^i \label{eqn:genseri} \ \ .
\end{equation}

In practical calculations one has only a finite number of coefficients
$c_i$, $0\leq i\leq n_c$ of the cycle expansion at disposal. Here $n_c$
is the cutoff in (topological) length. 
In \cite{PDJPAL97} we proposed a simple
resummation scheme for the computation of the coefficients $a_i$ and $b_i$
in (\ref{eqn:genseri}). 
We replace the infinite  in (\ref{eqn:genseri}) 
sums by finite sums
of increasing degrees, $n_a$ and $n_b$, and require that
\begin{equation}
\sum_{i=0}^{n_a} a_i (1-z)^i+
(1-z)^{1/s} \sum_{i=0}^{n_b} b_i
(1-z)^i=\sum_{n=0}^{n_c} c_n z^n+O(z^{n+1})
  \label{eqn:genserf} \ \ .
\end{equation}
One expands $(1-z)^{i(+1/s)}$ in binomial sums (series).
If $n_a +n_b+2=n+1$ this leads to a solvable linear system of equations
yielding the coefficients $a_i$ and $b_i$. 
It is natural to require that $|n_a +\frac{1}{s} -n_b|<1$ so that the maximal
powers of the two sums in eq. (\ref{eqn:genserf}) are adjacent.

If the zeta function is entire in the
entire $z$ plane (except for the branch cut) the periodic orbit sum 
can be written
\begin{equation}
Z_n=\sum_\alpha z_\alpha^{-n}
+\frac{1}{2\pi i} \int_{\gamma_{cut}}
z^{-n} 
\left( \frac{d}{dz} \log \zeta^{-1}(z) \right) dz  \  \  .
\label{eqn:periodic orbit sumsplit}
\end{equation}
The sum is over all zeroes $z_\alpha$ of the zeta function (assuming they are
not degenerated) and the contour $\gamma_{cut}$ goes round the
branch cut in positive direction.
If poles and/or natural boundaries are present, expression
(\ref{eqn:periodic orbit sumsplit}) must be accordingly modified.

The leading asymptotic behavior is provided by the
vicinity of the branch point $z=1$, and is found to be
\cite{DettDahl98}
\begin{equation}
Z_n \sim
\frac{b_0}{a_0} \frac{1}{s} \frac{1}{\Gamma(1-1/s)}\frac{1}{n^{1/s}}
\hspace{1cm} n \rightarrow \infty    \  \  .
\label{eqn:poweras}
\end{equation}

\begin{figure}
\centering
\leavevmode
\epsfxsize=10cm
\epsfbox{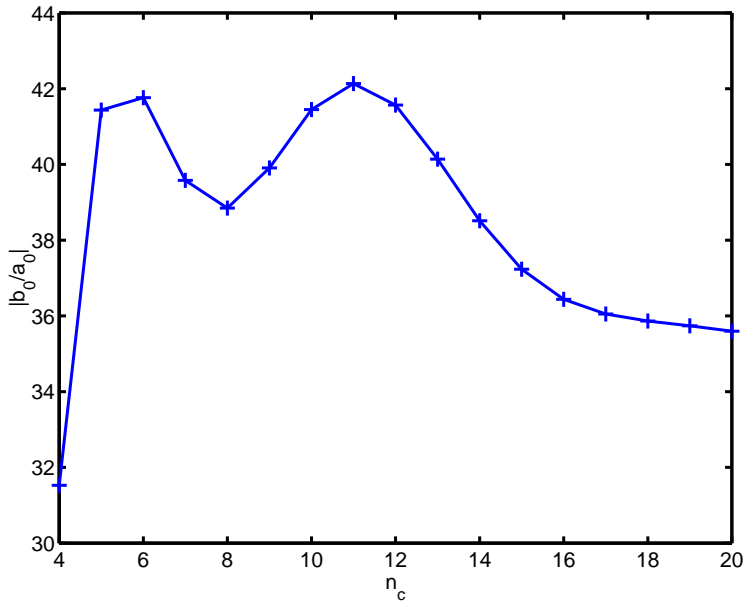}
\caption{The ratio $b_0/a_0$ versus cutoff length $n_c$.}
\label{fig:ratconv}
\end{figure}

The relevant
ratio $b_0/a_0$, obtained from the resummation scheme, versus cutoff length
$n_c$ is plotted in fig. \ref{fig:ratconv}.
In all numerical work we have used the parameters $s=0.7$ and $p=1.2$.
and computed all
periodic orbits up to length 20.

There is also a pair of complex conjugate zeroes, 
$z_0=x_0\pm i y_0$
close to the branch cut.
They contribute both to the sum over zeroes and to the integral around
the cut in (\ref{eqn:periodic orbit sumsplit}). But since their imaginary part $y_0$
is small,
they will, in effect, contribute a factor
$x_0^{-n}$ to the periodic orbit sum $Z_n$.

This zero will dominate $Z_n$ 
in some range $0\ll n \ll n_{cross}$
before the asymptotic power law sets in.

\begin{figure}
\centering
\leavevmode
\epsfxsize=10cm
\epsfbox{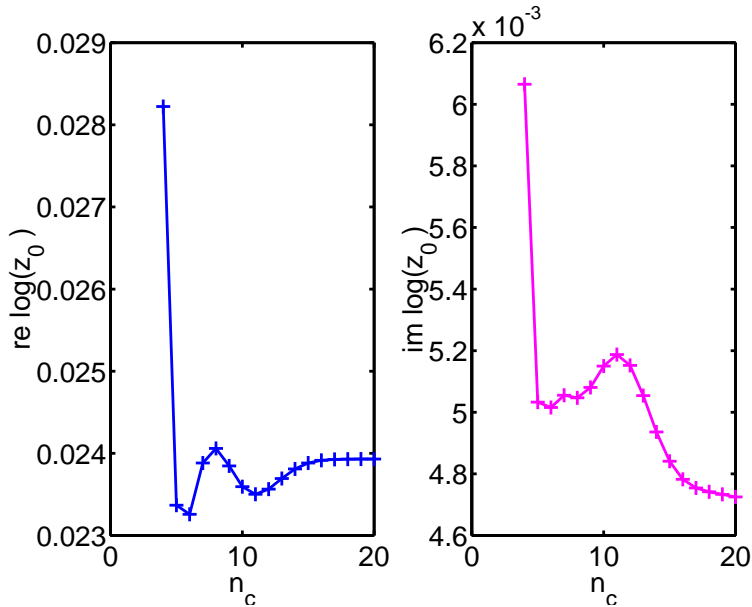}
\caption{The real and imaginary part of
$\log z_0$ versus cutoff length $n_c$.}
\label{fig:convergence}
\end{figure}

In fig. \ref{fig:convergence} we study the convergence of 
the real and imaginary part of $z_0$ obtained from the resummation
scheme above, for different cutoffs $n_c$.
The zero is computed by Newton-Raphson iteration of the
left hand side of (\ref{eqn:genserf}),
after the resummation has been done. Again we note that the analytic
continuation technique works quite satisfactorily.

The probability of escaping at iteration $n$ is
\begin{equation}
p_n=\Gamma_{n-1}-\Gamma_n  \  \   .
\label{eqn:pn}
\end{equation}
We get for this distribution
\begin{equation}
p_n \approx  \left\{
\begin{array}{ll}
x_0^{-n}  &   0\ll n \ll n_{cross}\\
\frac{b_0}{a_0} \frac{1}{s^2} 
\frac{1}{\Gamma(1-1/s)}\frac{1}{n^{1+1/s}} & n_{cross} \ll n 
\end{array}
\right.     \  \  .
 \label{eqn:cross}
\end{equation}
Here we have neglected the interval $I_{0^n}$, having the same 
asymptotic decay law as the periodic orbit sum $Z_n$. 
Due to the uncertainty in the bounds (\ref{eqn:boundhyper})
it can be neglected.

The crossover $n=n_{cross}$ 
takes place
when the two terms in (\ref{eqn:cross}) are of comparable
magnitude.
For our standard set of parameters
($p=1.2$, $s=0.7$) it is
found to be $n_{cross}\approx 300$.

\begin{figure}
\centering
\leavevmode
\epsfxsize=10cm
\epsfbox{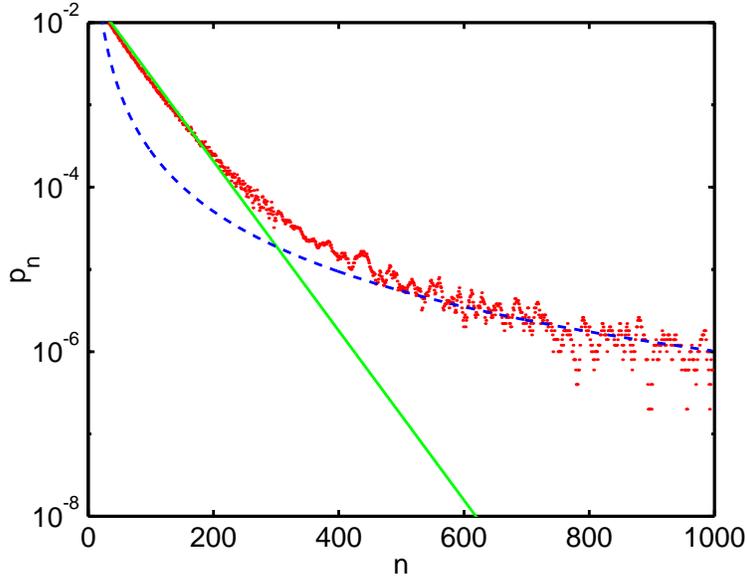}
\caption{Distribution of escape times obtained from
simulation (shaky curve) and the pre exponential (full line)
and the asymptotic power law (dashed line) obtained from resummation.}
\label{fig:simulation}
\end{figure}

The check our predictions we run a simulation of the system.
The result can be seen in fig.
(\ref{fig:simulation}). We note that the slope of the exponential,
the power and magnitude of the power law, as well as the crossover
time agrees very well with our predictions.

A reader still in any doubt on the effectiveness of cycle expansions
should consider the following.
The simulation  in fig. \ref{fig:simulation} 
averaged over $10^8$ initial point, yet, in itself the result would 
not very conclusive.
A direct evaluation of the periodic orbit sum up to say $n=600$ would require 
roughly $10^{170}$ periodic orbits. 
We have not bothered to perform such a cross check!
But a resummed cycle expansion
provides reliable answers with
a length cutoff as low as $n=15$, corresponding to $4719$ prime cycles.
Admittedly, we benefitted from knowing the asymptotic power law
of the cycle expansion. However, if this is not the situation, this power
law is easily extracted if one uses stability ordering \cite{DettDahl98}.

\section{A zero is cut in two}

The occurrence of a dominating zero beyond the branch point is, in fact, very
natural. Consider the one-parameter family of zeta functions.
\begin{equation}
\zeta^{-1}(z;\beta)=\prod_p \left( 1-\frac{z^{n_p}}{|\Lambda_p|^\beta}\right)
 \  \  .
\end{equation}
For $\beta$ small enough there is a leading zero $z_0(\beta)$ within
the domain of convergence $|z_0(\beta)|<1$. This is related to the topological pressure
\cite{Rue,Beck}
according to ${\cal P}(\beta)=-\log z_0(\beta)$.
For instance ${\cal P}(0)$ is the topological entropy.
For a certain $\beta$ (actually the fractal dimension of the repeller)
the zero collides with the branch point $z=1$, splits into two, and continue
to move out beyond the branch point. 

\begin{figure}
\centering
\leavevmode
\epsfxsize=10cm
\epsfbox{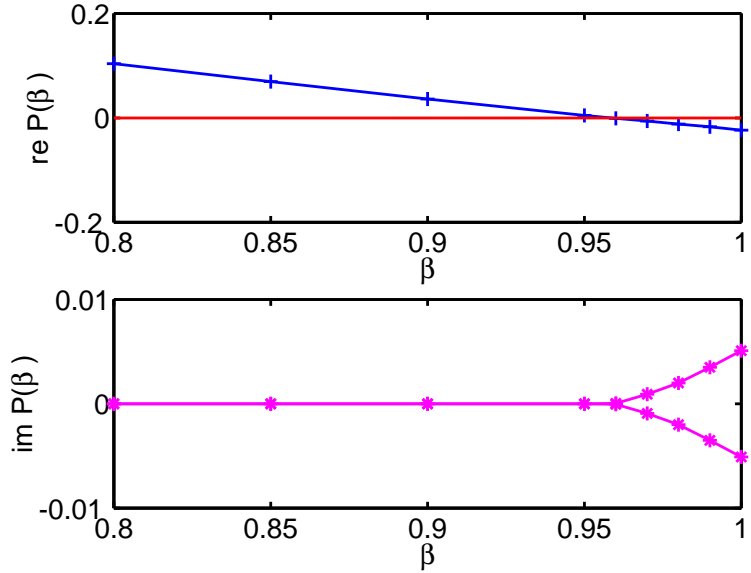}
\caption{The real and imaginary part of the leading zero 
versus the thermodynamic parameter $\beta$
}
\label{fig:pressure}
\end{figure}

In fig. \ref{fig:pressure} we plot the logarithm of the leading zero 
($-\log z_0(\beta)$)
versus $\beta$.
It is obtained from a resummation analogous to the one discussed above,
cf.\cite{PDJPAL97}.
It can be interpreted as the topological 
pressure only as long it is real.

\section{Mesoscopic discussion}

The particular form of the distribution of escape times
does depend on the initial distribution $\rho_0(x)$.
In this paper we have restricted ourselves
to a uniform initial distribution.
To model chaotic scattering one must imagine 
that particles can be injected according to
any distribution.
For example, one can construct a chaotic scatterer from a bounded billiard
by drilling holes wherever on the boundary and injecting particles from
different angles.
 This may even effect the asymptotic power law
\cite{Pik}. 
Periodic orbit theories can also account for other initial distributions
than uniform. However, the preceding discussions about relating
periodic orbit sums to survival probabilities warns us to be cautious
when doing so for intermittent systems.

As it appears,
the general rule of thumb, first an exponential, then a cross over to some 
powerlaw, can be extended
to open Hamiltonian systems with a mixed phase space structure \cite{Alt,Pik}.

An immediate application concerns conductance fluctuations in quantum dots
\cite{dots}. The Fourier transform $\hat{C}(x)$ of the 
correlation function $C(\Delta k)=\langle T(k)T(k+\Delta k) \rangle_k$,
where $T(k)$ is the transmission as function of the Fermi wave number,
can, after several 
approximations,
 be related to the escape distribution $p(L)$ \cite{dots}
\begin{equation}
\hat{C}(x) \sim \int_0^\infty dL p(L+x)p(L)  \   \   .
\end{equation}
If there is a crossover to a power law in $p(L)$ there will be an
associated crossover in $\hat{C}(x)$.
For an intermittent chaotic systems,
the crossover time may be very long - 
the quasi regular region component of phase space 
will not make itself noticed
until
very long times.
If the elastic mean free path of electrons 
is much shorter than the length corresponding to the crossover time, 
the quasi regular component will never ve detected in this type
of experiments.
Or the other way around, a small  deviation
from an integrable structure induces chaotic layers in phase space.
This chaotic layers may lead to exponential escape for small times,
and the experimental outcome may very well resemble
predictions for fully chaotic systems.

In experiments a (weak) magnetic field is a more natural control parameter
than the Fermi energy. 
Instead of the distribution of dwelling times one has to consider the
distribution of enclosed area, a related but more subtle concept
which we plan to address in future work.
One has observed Lorentzians shape
(predicted for chaotic systems ) of the so called weak localization peak
even in near integrable structure \cite{bird}. 
This has been attributed to naturally
occurring imperfections \cite{KF1,KF2} and rhymes well with the classical
considerations above.
Admittedly,
we have now moved far from our original intermittent map and entered
the realm of speculations.
What we do want to point out in this letter is that these kind of problem are 
well suited for periodic orbit computations - zeta functions is a 
powerful tool for
making long time predictions, even for intermittent chaos, once the problems
of
analytical continuation can be overcome.

\vspace{1cm}

I am grateful to Hans Henrik Rugh for pointing out an inconsistency,
in an
early version of this paper, and to Carl Dettmann for critical reading.
I would like to thank
Karl-Fredrik Berggren and
Igor Zozoulenko for interesting discussions.
This work was supported by the Swedish Natural Science
Research Council (NFR) under contract no. F-AA/FU 06420-314.

\newcommand{\PR}[1]{{Phys.\ Rep.}\/ {\bf #1}}
\newcommand{\PRL}[1]{{Phys.\ Rev.\ Lett.}\/ {\bf #1}}
\newcommand{\PRA}[1]{{Phys.\ Rev.\ A}\/ {\bf #1}}
\newcommand{\PRB}[1]{{Phys.\ Rev.\ B}\/ {\bf #1}}
\newcommand{\PRD}[1]{{Phys.\ Rev.\ D}\/ {\bf #1}}
\newcommand{\PRE}[1]{{Phys.\ Rev.\ E}\/ {\bf #1}}
\newcommand{\JPA}[1]{{J.\ Phys.\ A}\/ {\bf #1}}
\newcommand{\JPB}[1]{{J.\ Phys.\ B}\/ {\bf #1}}
\newcommand{\JCP}[1]{{J.\ Chem.\ Phys.}\/ {\bf #1}}
\newcommand{\JPC}[1]{{J.\ Phys.\ Chem.}\/ {\bf #1}}
\newcommand{\JMP}[1]{{J.\ Math.\ Phys.}\/ {\bf #1}}
\newcommand{\JSP}[1]{{J.\ Stat.\ Phys.}\/ {\bf #1}}
\newcommand{\AP}[1]{{Ann.\ Phys.}\/ {\bf #1}}
\newcommand{\PLB}[1]{{Phys.\ Lett.\ B}\/ {\bf #1}}
\newcommand{\PLA}[1]{{Phys.\ Lett.\ A}\/ {\bf #1}}
\newcommand{\PD}[1]{{Physica D}\/ {\bf #1}}
\newcommand{\NPB}[1]{{Nucl.\ Phys.\ B}\/ {\bf #1}}
\newcommand{\INCB}[1]{{Il Nuov.\ Cim.\ B}\/ {\bf #1}}
\newcommand{\JETP}[1]{{Sov.\ Phys.\ JETP}\/ {\bf #1}}
\newcommand{\JETPL}[1]{{JETP Lett.\ }\/ {\bf #1}}
\newcommand{\RMS}[1]{{Russ.\ Math.\ Surv.}\/ {\bf #1}}
\newcommand{\USSR}[1]{{Math.\ USSR.\ Sb.}\/ {\bf #1}}
\newcommand{\PST}[1]{{Phys.\ Scripta T}\/ {\bf #1}}
\newcommand{\CM}[1]{{Cont.\ Math.}\/ {\bf #1}}
\newcommand{\JMPA}[1]{{J.\ Math.\ Pure Appl.}\/ {\bf #1}}
\newcommand{\CMP}[1]{{Comm.\ Math.\ Phys.}\/ {\bf #1}}
\newcommand{\PRS}[1]{{Proc.\ R.\ Soc. Lond.\ A}\/ {\bf #1}}
%


\end{document}